\documentstyle[12pt,epsfig,cite]{article}

\def\be{\begin{equation}}
\def\ee{\end{equation}}
\def\ba{\begin{eqnarray}}
\def\ea{\end{eqnarray}}

\renewcommand{\thefootnote}{\fnsymbol{footnote}}
\begin{document}

\begin{titlepage}

\renewcommand{\thefootnote}{\fnsymbol{footnote}}

%\hfill SOGANG-MP 1/15

\vspace{0.3cm}

\begin{center}
{\Large\bf Generalized uncertainty principles and\\
black hole temperatures in rainbow gravity}
\end{center}

\begin{center}
Yong-Wan Kim\footnote{Electronic address:
ywkim65@gmail.com}$^{1}$, Seung Kook Kim\footnote{Electronic
address: skandjh@seonam.ac.kr}$^{2}$,  and Young-Jai
Park\footnote{Electronic address: yjpark@sogang.ac.kr}$^{3}$\par
\end{center}

\begin{center}

{${}^{1}$Research Institute of Physics and Chemistry,\\ Chonbuk
National University, Jeonju 54896, Korea,}\par {${}^{2}$Department
of Physical Therapy, Seonam University, Namwon 55724, Korea}\par
{${}^{3}$Department of Physics, Sogang University, Seoul 04107,
Korea}\par
\end{center}

\vskip 0.5cm
\begin{center}
{\today}
\end{center}

%\vfill
\vspace{0.5cm}

\begin{abstract}
In this paper, we have obtained modified black hole temperatures
in rainbow gravity by employing both the modified dispersion
relation (MDR) and the three different types of generalized
uncertainty principles (GUPs) including the extended uncertainty
principle (EUP). We also investigate their thermodynamic
stabilities of the modified Schwarzschild black hole according to
the different GUPs.

\end{abstract}

\vskip20pt

% PACS numbers: 04.70.Dy, 04.60.-m, 04.70.-s

% 04.70.Dy    Quantum aspects of black holes, evaporation, thermodynamics
% 04.60.-m    Quantum gravity
% 04.70.-s    Physics of black holes

%\vskip10pt


\end{titlepage}
%%%%%%%%%%%%%%%%%%%%%%%%%%%%%%%%%%%%%%%%%%%%%%%%%%%%%%%%%%%%%%%%%%%%%%%
\section{ Introduction}
\setcounter{equation}{0}
\renewcommand{\theequation}{\arabic{section}.\arabic{equation}}
%%%%%%%%%%%%%%%%%%%%%%%%%%%%%%%%%%%%%%%%%%%%%%%%%%%%%%%%%%%%%%%%%%%%%%

Hawking's semiclassical treatment~\cite{Hawking:1974sw} breaks
down as the size of the event horizon of a black hole approaches
the Planck length and quantum gravitational effects become
large~\cite{Maggiore:1993rv,Scardigli:1999jh}. One common feature
of quantum gravity effects is the existence of a minimal length
scale, which causes an alteration of the Heisenberg uncertainty
principle (HUP) to a generalized uncertainty principle (GUP)
\cite{Amati:1988tn,Konishi:1989wk,Kato:1990bd,Kempf:1993bq,Garay:1994en,Kempf:1996nk,
KalyanaRama:2001xd,Chang:2001bm,Medved:2004yu,Setare:2005sj,Nouicer:2005dp,Kim:2006rx,
Myung:2006qr,Kim:2007if,Nozari:2010qy,Ali:2011ap,Pedram:2011gw,Sprenger:2012uc,
Hossenfelder:2012jw,Dehghani:2016yjm,Zhou:2016cwy,Bernardo:2016dcy,Feng:2016tyt,
Tao:2017mpe,Kempf:1994su,Tawfik:2014zca,Anacleto:2015rlz,Bolen:2004sq,Bambi:2007ty,Park:2007az}.
A GUP again significantly modifies the Hawking radiation and
leaves a remnant at the final stage of the
evaporation~\cite{Adler:2001vs}. Moreover, the existence of a
minimal length scale makes us puzzled since there is no length
invariant under the special relativity. This contemplation leads
to doubly special relativity (DSR), or special relativity with two
observer independent
scales~\cite{AmelinoCamelia:2000ge,AmelinoCamelia:2000mn,AmelinoCamelia:2002wr,
KowalskiGlikman:2004qa}. Of course, it is assumed that the DSR
becomes the standard special relativity in the limit of vanishing
the minimal length. It is naturally expected in the DSR that the
usual energy-momentum dispersion relation should be modified by
some nonlinear mass-shell relations.

On the other hand, Magueijo and Smolin (MS) \cite{Magueijo:2002xx}
have extended the DSR to general relativity by proposing that
quanta of different energies see different background geometry,
referred to as rainbow gravity. Since then many efforts have been
devoted to rainbow gravity related to extending gravity and other
stimulated work at the Planck scale
\cite{Liberati:2004ju,Galan:2004st,Galan:2005ju,
Hackett:2005mb,Ling:2006az,
Ling:2006ba,Girelli:2006fw,Li:2008gs,Ling:2008sy,
Garattini:2011kp,Garattini:2011fs,Garattini:2012ec,Garattini:2012ca,Majumder:2013mza,
Amelino-Camelia:2013wha,Awad:2013nxa,Barrow:2013gia,Santos:2015sva,
Carvalho:2015omv,Ashour:2016cay,Ali:2014zea,Gim:2015zra,Hendi:2015cra,
Kim:2015wwa,Gim:2015yxa,Hendi:1512,Hendi:2016vux,Tao:2016baz,Yadav:2016nfh,Gangopadhyay:2016rpl,
Hendi:2016njy,Banerjee:2016zcu,Hendi:2016hbe,Momeni:2017cvl,Alsaleh:2017oae,
Feng:1708,Deng:2017umx}. Recently, Ali~\cite{Ali:2014xqa}
calculated thermodynamic quantities in rainbow Schwarzschild black
hole with a particular choice of rainbow functions and studied
corresponding thermodynamics. Here, he obtained the black hole
temperature from the definition of the surface gravity in the
rainbow Schwarzschild black hole and then by invoking the HUP and
the ordinary dispersion relation the energy dependence in the
temperature was eliminated. Later, Gim and Kim~\cite{Gim:2014ira}
improved the black hole temperature whose energy dependence was
eliminated by employing both the HUP and the MDR. Very recently,
we have extended the previous study of the thermodynamics and
phase transition of the rainbow Schwarzschild black hole to the
Schwarzschild-AdS black hole~\cite{Kim:2016qtp} where metric
depends on the energy of a probe by employing both the HUP and the
MDR.

In this paper, we will further investigate modified temperatures
of black holes in rainbow gravity by fully employing both the MDR
and the various GUPs, not just restricted by the HUP, and study
their corresponding thermodynamic stabilities. In section 2, we
first briefly review the method of Adler-Chen-Santiago (ACS)
heuristically to find black hole's temperatures related to the
HUP. Then, we obtain the desired black hole's temperature
considering three different types of the GUPs including the
extended uncertainty principle (EUP) and the generalized extended
uncertainty principle (GEUP). In section 3, we study the GUPs
effects in black hole temperatures making full use of the MDR in
rainbow gravity. In section 4, we investigate thermodynamic
stabilities of both the GUP and the MDR corrected Schwarzschild
black holes in rainbow gravity. Finally, a conclusion and
discussion is given in section 5.

%%%%%%%%%%%%%%%%%%%%%%%%%%%%%%%%%%%%%%%%%%%%%%%%%%%%%%%%%%%%%%%%%%%%%%%
\section{Black hole temperatures with various GUPs}
\setcounter{equation}{0}
\renewcommand{\theequation}{\arabic{section}.\arabic{equation}}
%%%%%%%%%%%%%%%%%%%%%%%%%%%%%%%%%%%%%%%%%%%%%%%%%%%%%%%%%%%%%%%%%%%%%%%
%%%%%%%%%%%%%%%%%%%%%%%%%%%%%%%%%%%%%%%%%%%%%%%%%%%%%%%%%%%%%%%%%%%%%%%
%\subsection{HUP}
%%%%%%%%%%%%%%%%%%%%%%%%%%%%%%%%%%%%%%%%%%%%%%%%%%%%%%%%%%%%%%%%%%%%%%%

According to ACS \cite{Adler:2001vs}, let us first briefly
recapitulate the way of getting the well-known Hawking temperature
by considering the HUP for a spherically symmetric black hole.

In the vicinity of a black hole surface, there is intrinsic
position uncertainty of the Hawking radiated photons of order of
the event horizon $r_+$ as
\begin{equation}
\Delta x \sim r_+.
\end{equation}
On the other hand, according to the HUP, it leads to momentum
uncertainty of order $p$ as
\begin{equation}\label{hup-mom-uncertaity}
p=\Delta p\sim \frac{\hbar}{\Delta x}\sim\frac{\hbar}{r_+}.
\end{equation}
Let us plug this momentum uncertainty into the standard dispersion
relation
\begin{equation}\label{DR}
E^2-p^2c^2=m^2.
\end{equation}
Then, for a massless particle, the energy uncertainty will be
\begin{equation}
E = p c \sim \frac{\hbar c}{r_+},
\end{equation}
which can be identified with the characteristic energy of the
emitted photon as $E=k_B T$. As a result, it concludes that a
black hole's temperature be
\begin{equation}
T^{HUP}=\frac{1}{4\pi r_+}
\end{equation}
with a calibration factor of $1/4\pi$ and $k_B=\hbar=c=1$
\footnote[2]{Hereafter, we will use the natural units of $\hbar=1$
and $c=1$ except in the uncertainty principles themselves for
their clarity.}, which is in Fig. 1(a). The figure shows that the
temperature diverges as $r_+\rightarrow 0$, while it vanishes as
$r_+\rightarrow\infty$. Note that for the case of the
Schwarzschild black hole with $r_+=2GM$, it gives exactly the same
Hawking temperature as
\begin{equation}
T^{HUP}_{H}=\frac{1}{8\pi GM}.
\end{equation}

%%%%%%%%%%%%%%%%%%%%%%%%%%%%%%%%%%%%%%%%%%%%%%%%%%%%%%%%%%%%%%%%%%%%%%%
\subsection{GUP case}
%%%%%%%%%%%%%%%%%%%%%%%%%%%%%%%%%%%%%%%%%%%%%%%%%%%%%%%%%%%%%%%%%%%%%%%

In order to extend the method of ACS to various GUPs, let us first
consider the GUP case which is susceptible when the momentum is
close to the Planck scale, given by the inequality
\begin{equation}
 \Delta x \geq \frac{\hbar}{\Delta p}+\alpha\ell^2_p\frac{\Delta
 p}{\hbar},
\end{equation}
where $\alpha$ is a dimensionless constant and
$\ell^2_p=G\hbar/c^3$ is the Planck length
\cite{Maggiore:1993rv,Scardigli:1999jh,Amati:1988tn,Konishi:1989wk,Kato:1990bd,
Kempf:1993bq,Garay:1994en,Kempf:1996nk,
KalyanaRama:2001xd,Chang:2001bm,Medved:2004yu,Setare:2005sj,Nouicer:2005dp,Kim:2006rx,
Myung:2006qr,Kim:2007if,Nozari:2010qy,Ali:2011ap,Pedram:2011gw,Sprenger:2012uc,
Hossenfelder:2012jw,Dehghani:2016yjm,Zhou:2016cwy,Bernardo:2016dcy,Feng:2016tyt,
Tao:2017mpe,Kempf:1994su,Tawfik:2014zca,Anacleto:2015rlz,Bolen:2004sq,Park:2007az}.
This gives us the momentum uncertainty of
\begin{equation}\label{gupreln}
 \frac{\Delta x}{2\alpha\ell^2_p}\left(1-\sqrt{1-\frac{4\alpha\ell^2_p}{(\Delta x)^2}}\right)
 \leq
 \frac{\Delta p}{\hbar}
 \leq\frac{\Delta x}{2\alpha\ell^2_p}\left(1+\sqrt{1-\frac{4\alpha\ell^2_p}{(\Delta x)^2}}\right).
\end{equation}
It explicitly shows that there exists an absolute minimum in the
position uncertainty as
\begin{equation}\label{min-uncertainty}
\Delta x \geq \Delta x_{\rm min} \equiv 2\sqrt{\alpha}\ell_p.
\end{equation}
The existence of $\Delta x_{\rm min}$ implies that the position
uncertainty cannot be arbitrarily small, while the momentum
uncertainty cannot be arbitrarily large. Note that as
$\alpha\rightarrow 0$, or, equivalently, when $\Delta x \gg \Delta
x_{\rm min}$ in the semiclassical regime, the HUP can be recovered
from the left inequality in Eq. (\ref{gupreln}).

Now, according to ACS \cite{Adler:2001vs} as before, the momentum
uncertainty of order $p$ can be obtained from the position
uncertainty of an emitted photon of order   $\Delta x \sim r_+$ as
\begin{equation}\label{GUP-p}
p=\Delta p
\sim\frac{r_+}{2\alpha\ell^2_p}\left(1-\sqrt{1-\frac{4\alpha\ell^2_p}{r^2_+}}\right).
\end{equation}
Plugging this into the standard dispersion relation for a massless
particle, the energy uncertainty gives a GUP--corrected black hole
temperature as
\begin{equation}\label{GUP-Temp}
T^{GUP}=\frac{r_+\omega^2_p}{8\pi\alpha}\left(1-\sqrt{1-\frac{4\alpha}{r^2_+\omega^2_p}}\right)
\end{equation}
with the calibration factor of $1/4\pi$. This GUP-modified
temperature is plotted in Fig. 1(b), which shows that the
temperature vanishes as $r_+\rightarrow\infty$. On the other hand,
since $r_+$ cannot go through the minimum position uncertainty
$\Delta x_{\rm min}$ as like in Eq.~(\ref{min-uncertainty}), the
evaporation of the GUP-modified temperature stops at
$r_{+,0}=2\sqrt{\alpha}/\omega_p$ where
$T^{GUP}_0=\omega_p/4\pi\sqrt{\alpha}$, leaving a remnant. Note
that here we have used the left inequality in Eq.~(\ref{gupreln})
for the momentum uncertainty since it recovers correctly the HUP
limit in the semiclassical regime, and $\omega_p=\ell^{-1}_p$. And
for the Schwarzschild black hole with $r_+=2GM$, the GUP-modified
Hawking temperature is reduced to
\begin{equation}
T^{GUP}_H=\frac{M}{4\pi}\left(1-\sqrt{1-\frac{\omega^2_p}{M^2}}\right)
\end{equation}
with $G=\omega^{-2}_p$, which shows that the total Hawking
evaporation of the black hole is prevented by the effect of the
minimal length in the GUP.

\begin{figure*}[t!]
   \centering
   \includegraphics{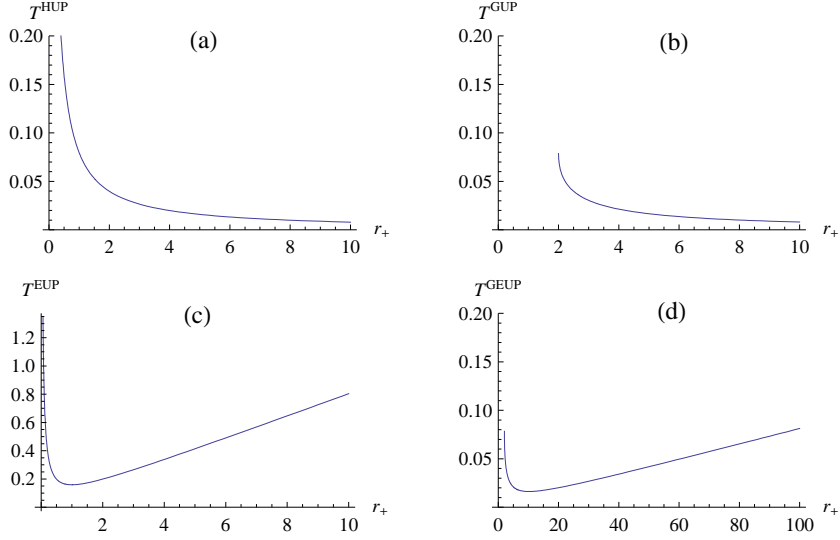}
\caption{Hawking temperatures using (a) the HUP, (b) the GUP, (c)
the EUP, and (d) the GEUP.}
 \label{fig1}
\end{figure*}

%%%%%%%%%%%%%%%%%%%%%%%%%%%%%%%%%%%%%%%%%%%%%%%%%%%%%%%%%%%%%%%%%%%%%%%
\subsection{EUP case}
%%%%%%%%%%%%%%%%%%%%%%%%%%%%%%%%%%%%%%%%%%%%%%%%%%%%%%%%%%%%%%%%%%%%%%%

Second, the extended uncertainty principle (EUP) is given by
\begin{equation}
\Delta p \geq \frac{\hbar}{\Delta
x}+\frac{\hbar\beta^2}{\ell^2}\Delta x,
\end{equation}
where $\beta$ is a dimensionless constant and $\ell$ the curvature
radius of the AdS spacetime
\cite{Bolen:2004sq,Bambi:2007ty,Park:2007az,Li:2008gs}. This EUP
shows that the momentum uncertainty cannot be arbitrarily small
but $\Delta p\geq \Delta p_{\rm min}\equiv 2\hbar\beta^2/\ell^2$.

According to ACS \cite{Adler:2001vs} as before, the momentum
uncertainty of order $p$ can also be obtained from a position
uncertainty of an emitted photon of order of $\Delta x \sim r_+$
as
\begin{equation}\label{EUP-p}
p=\Delta
p\sim\frac{1}{r_+}\left(1+\frac{\beta^2}{\ell^2}r^2_+\right).
\end{equation}
This momentum uncertainty with the standard dispersion relation
for a massless particle gives a EUP-corrected black hole
temperature as
\begin{equation}\label{EUP-Temp}
T^{EUP}=\frac{1}{4\pi
r_+}\left(1+\frac{\beta^2}{\ell^2}r^2_+\right)
\end{equation}
with the same calibration factor as before. The modified
temperature obtained using the EUP was plotted in Fig. 1(c), which
shows that it behaves as like the case of the Schwarzschild-AdS
black hole; it is divergent not only as $r_+\rightarrow\infty$ but
also as $r_+\rightarrow 0$. A global minimum temperature,
$T^{EUP}_{\rm min}=\beta/2\pi\ell$, is achieved at $r_{+,~{\rm
min}}=\ell/\beta$. In addition, for $\beta^2=3$, it is exactly the
same with the Hawking temperature of the Schwarzschhild-AdS black
hole so that one sees that in the presence of the cosmological
constant the EUP leads to the Hawking temperature for the
Schwarzschhild-AdS black hole. Note also that replacing $\ell^2$
with $-\ell^2$ one can obtain the Hawking temperature of the
Schwarzschhild--dS black hole.

%%%%%%%%%%%%%%%%%%%%%%%%%%%%%%%%%%%%%%%%%%%%%%%%%%%%%%%%%%%%%%%%%%%%%%%
\subsection{GEUP case}
%%%%%%%%%%%%%%%%%%%%%%%%%%%%%%%%%%%%%%%%%%%%%%%%%%%%%%%%%%%%%%%%%%%%%%%

Finally, the generalized extended uncertainty principle (GEUP)
\cite{Kempf:1994su,Tawfik:2014zca,Anacleto:2015rlz,Bolen:2004sq,Bambi:2007ty,Park:2007az}
is described by
\begin{equation}
\Delta x \Delta p \geq \hbar+\alpha\ell^2_p\frac{(\Delta
p)^2}{\hbar} +\frac{\hbar\beta^2}{\ell^2}(\Delta x)^2.
\end{equation}
This gives us the momentum uncertainty of
\begin{eqnarray}
 && \frac{\Delta x}{2\alpha\ell^2_p}
 \left[1-\sqrt{1-\frac{4\alpha\ell^2_p}{(\Delta x)^2}\left(1+\frac{\beta^2}{\ell^2}(\Delta x)^2\right)}\right]
 \leq
 \frac{\Delta p}{\hbar}\leq \nonumber\\
 &&~~~
  \frac{\Delta x}{2\alpha\ell^2_p}
 \left[1+\sqrt{1-\frac{4\alpha\ell^2_p}{(\Delta x)^2}\left(1+\frac{\beta^2}{\ell^2}(\Delta x)^2\right)}\right].
\end{eqnarray}
It also says that the position uncertainty is bounded from below
by
\begin{equation}
 \Delta x \ge \Delta x_{\rm min}\equiv
 \frac{2\sqrt{\alpha}\ell_p}{\sqrt{1-4\alpha\beta^2\ell^2_p/\ell^2}}.
\end{equation}

Now, according to ACS \cite{Adler:2001vs} as before, the momentum
uncertainty of order $p$ can be obtained from a position
uncertainty of an emitted photon of order of $\Delta x \sim r_+$
as
\begin{equation}\label{GEUP-p}
p=\Delta p\sim \frac{r_+}{2\alpha\ell^2_p}
 \left[1-\sqrt{1-\frac{4\alpha\ell^2_p}{r^2_+}\left(1+\frac{\beta^2}{\ell^2}r^2_+\right)}\right].
\end{equation}
Plugging this momentum uncertainty into the standard dispersion
relation for a massless particle, the energy uncertainty gives a
GEUP-corrected black hole temperature as
\begin{equation}
T^{GEUP}=\frac{r_+\omega^2_p}{8\pi\alpha}
 \left[1-\sqrt{1-\frac{4\alpha\ell^2_p}{r^2_+}\left(1+\frac{\beta^2}{\ell^2}r^2_+\right)}\right]
\end{equation}
with the calibration factor of $1/4\pi$. In Fig. 1(d), we have
plotted the GEUP-corrected black hole temperature which shows a
mixture of the results of the GUP-corrected and the EUP-corrected
black hole temperatures; the evaporation of the GEUP-corrected
black hole temperature stops at the minimum position uncertainty
$\Delta x_{\rm min}=r_{+,0}$ as
\begin{equation}
 T^{GEUP}_0=\frac{\omega_p}{4\pi\sqrt{\alpha}}\frac{1}{\sqrt{1-4\alpha\beta^2/\ell^2\omega^2_p}},
\end{equation}
where
\begin{equation}
r_{+,0}=\frac{2\sqrt{\alpha}}{\omega_p}\frac{1}{\sqrt{1-4\alpha\beta^2/\ell^2\omega^2_p}},
\end{equation}
while the GEUP-corrected black hole temperature diverges as
$r_+\rightarrow\infty$.

Note that in the GUP limit of $\beta\rightarrow 0$, the
GEUP-corrected black hole temperature becomes
\begin{equation}
 T^{GEUP}\approx
 T^{GUP}+\beta\left(\frac{r_+}{4\pi\ell\sqrt{1-4\alpha/r^2_+\omega^2_p}}\right)
  +{\cal O}(\beta^2).
\end{equation}
On the other hand, in the EUP limit of $\alpha\rightarrow 0$, it
also well recovers the EUP-corrected black hole temperature as
\begin{equation}
 T^{GEUP}\approx
 T^{EUP}+\alpha\left(\frac{(1+\beta^2 r^2_+/\ell^2)^2}{4\pi r^3_+\omega^2_p}\right)
  +{\cal O}(\alpha^2).
\end{equation}

%%%%%%%%%%%%%%%%%%%%%%%%%%%%%%%%%%%%%%%%%%%%%%%%%%%%%%%%%%%%%%%%%%%%%%%
\section{GUPs Effects in rainbow gravity}
\setcounter{equation}{0}
\renewcommand{\theequation}{\arabic{section}.\arabic{equation}}
%%%%%%%%%%%%%%%%%%%%%%%%%%%%%%%%%%%%%%%%%%%%%%%%%%%%%%%%%%%%%%%%%%%%%%%

One of the drastic changes near the Planck scale one may think is
the modification of the standard dispersion relation (\ref{DR})
which was suggested in rainbow gravity as
\begin{equation}\label{MDR}
\omega^2f^2(\omega/\omega_p)-p^2c^2g^2(\omega/\omega_p)=m^2,
\end{equation}
where $f(\omega/\omega_p)$, $g(\omega/\omega_p)$ are called the
rainbow functions which satisfy with the conditions of
$\lim_{\omega \rightarrow 0} f(\omega/\omega_p) = 1$ and
$\lim_{\omega \rightarrow 0} g(\omega/\omega_p) = 1$ at low
energies. The MDR (\ref{MDR}), which is invariant under nonlinear
Lorentz transformation in the momentum space, can be realized in
the dual position space as one parameter energy dependent family
of metric \cite{Magueijo:2002xx} as
\begin{equation}
\label{rainbow-metric}
 ds^2=-\frac{F(r)}{f^2(\omega/\omega_p)}dt^2
      +\frac{G(r)}{g^2(\omega/\omega_p)}dr^2
      +\frac{r^2}{g^2(\omega/\omega_p)}d\Omega^2
\end{equation}
for a spherically symmetric spacetime. One of the most interesting
rainbow functions are described by
\begin{equation}\label{rainbowfunc}
f(\omega/\omega_p)=1, \quad g(\omega/\omega_p)=\sqrt{1-\eta
\left({\omega}/{\omega_p}\right)^n},
\end{equation}
which forms have been studied much in
Ref.\cite{Ali:2014zea,Gim:2015zra,Hendi:2015cra,Kim:2015wwa,Gim:2015yxa,
Gangopadhyay:2016rpl,Ali:2014xqa,Gim:2014ira,Kim:2016qtp,Hendi:1512,
Ashour:2016cay,Tao:2016baz,Hendi:2016njy,Banerjee:2016zcu,Momeni:2017cvl,
Alsaleh:2017oae} including the quantum gravity phenomenology
\cite{AmelinoCamelia:2008qg} and the $\kappa$-Minkowski
noncommutative spacetime \cite{Lukierski:1993wx}.  Note that the
choice of $f(\omega/\omega_p) = 1$ in the rainbow functions makes
a time-like Killing vector in the rainbow gravity as usual and so
the local thermodynamic energy independent of a test particle's
energy. Moreover, we have recently shown that a freely falling
observer sees his local temperature dependent only on
$g(\omega/\omega_p)$, needless to think about $f(\omega/\omega_p)$
\cite{Kim:2015wwa}. In the subsequent analysis, we will
particularly choose $n=2$ in the rainbow function of
$g(\omega/\omega_p)$ without loss of generality.

%%%%%%%%%%%%%%%%%%%%%%%%%%%%%%%%%%%%%%%%%%%%%%%%%%%%%%%%%%%%%%%%%%%%%%%
%\subsection{HUP in rainbow gravity}
%%%%%%%%%%%%%%%%%%%%%%%%%%%%%%%%%%%%%%%%%%%%%%%%%%%%%%%%%%%%%%%%%%%%%%%

Now, having seen the efficient way of ACS with the standard
dispersion relation in deriving Hawking temperatures in the
previous section, we will generalize the method to incorporate the
MDR (\ref{MDR}) in the rainbow gravity to derive Hawking
temperatures in this section.

Let us first briefly summarize the HUP result in the rainbow
gravity. As like in the section 2, one can relate the momentum
uncertainty of order $p$ with the position uncertainty of an
emitted photon of order of $\Delta x \sim r_+$ as in Eq.
(\ref{hup-mom-uncertaity}) using the HUP. Then, plugging this
momentum uncertainty into the MDR for a massless particle, one can
have
\begin{equation}\label{MDR-rainbow}
 E=pc\frac{g(\omega/\omega_p)}{f(\omega/\omega_p)}
  \sim \frac{\hbar c}{r_+}\frac{g(\omega/\omega_p)}{f(\omega/\omega_p)}.
\end{equation}
Therefore, one can directly obtain a black hole's temperature in
the rainbow gravity as
\begin{eqnarray}
\tilde{T}^{HUP}
 &=& \frac{1}{4\pi r_+}\frac{g(\omega/\omega_p)}{f(\omega/\omega_p)}\nonumber\\
 &=& T^{HUP}\frac{g(\omega/\omega_p)}{f(\omega/\omega_p)}
\end{eqnarray}
with the calibration factor of $1/4\pi$. This shows the effect of
the rainbow gravity permeates the black hole's temperature through
the rainbow functions.

When used the rainbow functions (\ref{rainbowfunc}) with $n=2$,
one has the black hole's temperature explicitly as
\begin{eqnarray}
 \tilde{T}^{HUP} &=& \frac{1}{4\pi r_+}
          \sqrt{1-\eta
          \left(\frac{\omega}{\omega_p}\right)^2}\nonumber\\
          &=& T^{HUP}
          \sqrt{1-\eta \left(\frac{\omega}{\omega_p}\right)^2}.
\end{eqnarray}
Now, by making use of the HUP (\ref{hup-mom-uncertaity}) and MDR
(\ref{MDR-rainbow}), one can eliminate the energy $\omega$
dependence and obtain the HUP-corrected black hole temperature in
the rainbow gravity as
\begin{equation}\label{hup-corr-T}
 \tilde{T}^{HUP} =  \frac{1}{4\pi r_+}
          \frac{r_+\omega_p}{\sqrt{r^2_+\omega^2_p+\eta}},
\end{equation}
which becomes the usual Hawking temperature as
$\omega_p\rightarrow\infty$. In Fig. 2(a), we have plotted the
HUP-corrected black hole temperature in the rainbow gravity. This
extends the previous results \cite{Ali:2014xqa,Gim:2014ira} where
the modified temperature is obtained from the HUP and the standard
dispersion relation to the one using the HUP and the MDR. As a
result, by considering the rainbow gravity, one can see that the
modified black hole temperature has no divergence at $r_+=0$, but
finite, $\tilde{T}^{HUP}_0=\omega_p/4\pi\sqrt{\eta}$, so that
leaves a remnant.

\begin{figure*}[t!]
   \centering
   \includegraphics{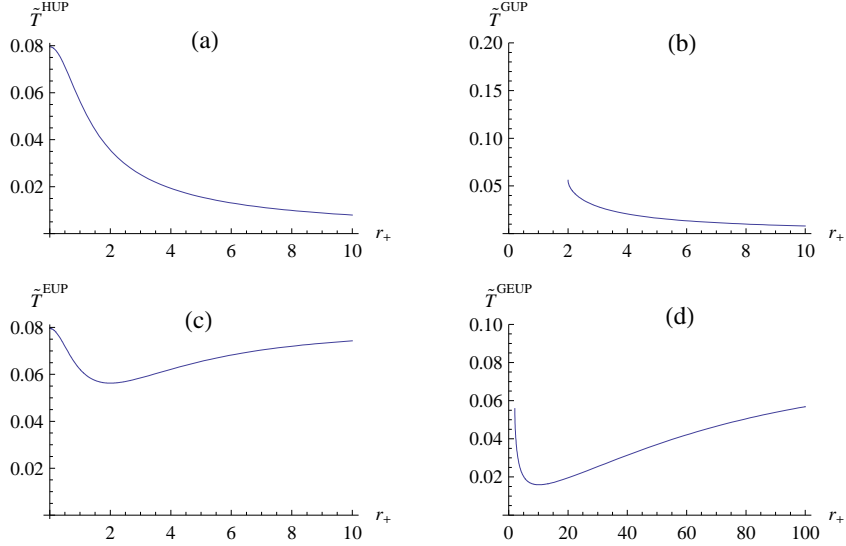}
\caption{Hawking temperatures in the rainbow gravity using (a) the
HUP, (b) the GUP, (c) the EUP, and (d) the GEUP.}
 \label{fig2}
\end{figure*}

%%%%%%%%%%%%%%%%%%%%%%%%%%%%%%%%%%%%%%%%%%%%%%%%%%%%%%%%%%%%%%%%%%%%%%%
\subsection{GUP in rainbow gravity}
%%%%%%%%%%%%%%%%%%%%%%%%%%%%%%%%%%%%%%%%%%%%%%%%%%%%%%%%%%%%%%%%%%%%%%%

Now, we can extend the previous analysis finding a black hole's
temperature to include the various GUPs.

First, by combining the momentum uncertainty (\ref{GUP-p})obtained
from a position uncertainty of an emitted photon of order of
$\Delta x \sim r_+$ in the GUP case with the MDR (\ref{MDR}) and
the rainbow functions (\ref{rainbowfunc}) for a massless particle,
one can obtain the GUP-corrected black hole temperature as
\begin{eqnarray}\label{TGUP-rainbow}
 \tilde{T}^{GUP} &=&
 T^{GUP}\frac{g(\omega/\omega_p)}{f(\omega/\omega_p)}\nonumber\\
 &=& T^{GUP} \sqrt{1-\eta \left(\frac{\omega}{\omega_p}\right)^2}.
\end{eqnarray}
Next, by making use of the momentum uncertainty (\ref{GUP-p}) in
the GUP case and MDR (\ref{MDR-rainbow}), one can easily find the
ratio of
\begin{equation}
\frac{\omega^2}{\omega^2_p}=\frac{\frac{r^2_+\omega^2_p}{2}\left(1-\sqrt{1-\frac{4\alpha}{r^2_+\omega^2_p}}\right)-\alpha}
    {\alpha^2+\eta\left(\frac{r^2_+\omega^2_p}{2}\left(1-\sqrt{1-\frac{4\alpha}{r^2_+\omega^2_p}}\right)-\alpha\right)}.
 \end{equation}
This can be used to eliminate the energy $\omega$ dependence in
the GUP-corrected black hole's temperature (\ref{TGUP-rainbow}) in
the rainbow gravity, and one can finally have
\begin{equation}\label{GUP-HT}
 \tilde{T}^{GUP} = T^{GUP}\left\{1-\eta
  \left[\frac{\frac{r^2_+\omega^2_p}{2}\left(1-\sqrt{1-\frac{4\alpha}{r^2_+\omega^2_p}}\right)-\alpha}
    {\alpha^2+\eta\left(\frac{r^2_+\omega^2_p}{2}\left(1-\sqrt{1-\frac{4\alpha}{r^2_+\omega^2_p}}\right)-\alpha\right)}
  \right]
                       \right\}^{\frac{1}{2}}.
\end{equation}
In Fig. 2(b), we have plotted the GUP-corrected black hole's
temperature in the rainbow gravity which as like the temperature
$T^{GUP}_0$ in Eq.~(\ref{GUP-Temp}) without considering the
rainbow gravity, it stops to evaporate at $r_{+,0}=\Delta x_{\rm
min}=2\sqrt{\alpha}/\omega_p$ where
$\tilde{T}^{GUP}_0=\omega_p/4\pi\sqrt{\alpha+\eta}$, which is
lower than $T^{GUP}_0$. And as $r_+\rightarrow\infty$, the
temperature $\tilde{T}^{GUP}$ approaches the usual Hawking
temperature of $T^{HUP}$ disappearing the strong gravity effect.
Moreover, when the rainbow effect is turned off, {\it i.e.},
$\eta\rightarrow 0$, the modified temperature becomes the
GUP-corrected black hole temperature (\ref{GUP-Temp}), and when
the GUP effect is turned off, {\it i.e.}, $\alpha\rightarrow 0$,
it becomes the HUP-corrected black hole temperature
(\ref{hup-corr-T}).

It is appropriate to comment that the GUP-corrected black hole's
temperature (\ref{GUP-HT}) is obtained from the consideration of
both the MDR and the GUP along with the definition of the
temperature in the rainbow gravity as like in Eq.
(\ref{TGUP-rainbow}). On the other hand, the similar result can be
found in Ref.~\cite{Gim:2014ira} in the conclusion. However, the
modified temperature discussed in there was obtained from the
consideration of the MDR and the GUP, but not considered the
rainbow effect in the temperature, which means that the authors
have used the standard dispersion relation to define the
temperature.

%%%%%%%%%%%%%%%%%%%%%%%%%%%%%%%%%%%%%%%%%%%%%%%%%%%%%%%%%%%%%%%%%%%%%%%
\subsection{EUP in rainbow gravity}
%%%%%%%%%%%%%%%%%%%%%%%%%%%%%%%%%%%%%%%%%%%%%%%%%%%%%%%%%%%%%%%%%%%%%%%

Second, by combining the momentum uncertainty (\ref{EUP-p})
obtained from a position uncertainty of an emitted photon of order
of $\Delta x \sim r_+$ in the EUP case with the MDR
(\ref{MDR-rainbow}) and the rainbow functions (\ref{rainbowfunc})
for a massless particle, one can obtain the EUP-corrected black
hole temperature in the rainbow gravity as
\begin{eqnarray}
 \tilde{T}^{EUP} &=&
 T^{EUP}\frac{g(\omega/\omega_p)}{f(\omega/\omega_p)}\nonumber\\
                 &=&  T^{EUP} \sqrt{1-\eta \left(\frac{\omega}{\omega_p}\right)^2}.
\end{eqnarray}
Moreover, one can determine the ratio of the energy
$\omega^2/\omega^2_p$ in the square root by making use of the
momentum uncertainty (\ref{EUP-p}) and the MDR (\ref{MDR-rainbow})
for the massless particle, and finally obtain the EUP-corrected
black hole temperature in the rainbow gravity as
\begin{equation}\label{EUP-HT}
 \tilde{T}^{EUP} = T^{EUP}\left\{1-\eta
  \left[\frac{{\cal B(\beta)}^2}
    {r^2_+\omega^2_p+\eta{\cal B(\beta)}^2}
  \right]
                       \right\}^{\frac{1}{2}},
\end{equation}
where ${\cal B(\beta)}\equiv 1+\frac{\beta^2}{\ell^2}r^2_+$. We
have plotted the EUP-corrected black hole temperature in the
rainbow gravity in Fig. 2(c). Comparing with the temperature
(\ref{EUP-Temp}) without the rainbow effect, the EUP-corrected
black hole temperature in the rainbow gravity is finite both at
$r_+=0$ and as $r_+\rightarrow\infty$, having and approaching
$\tilde{T}^{EUP}_0=\omega_p/4\pi\sqrt{\eta}$. It also has a global
minimum temperature,
$\tilde{T}^{EUP}_m=\omega_p/4\pi\sqrt{\eta+\ell^2\omega^2_p/4\beta^2}$,
at $r_+=r_m=\ell/\beta$.

On the other hand, when the rainbow effect disappears with
$\eta\rightarrow 0$, the modified temperature becomes the
EUP-corrected black hole temperature (\ref{EUP-Temp}), and when
the EUP effect is turned off with $\beta\rightarrow 0$, it
reproduces the HUP-corrected black hole temperature
(\ref{hup-corr-T}).

%%%%%%%%%%%%%%%%%%%%%%%%%%%%%%%%%%%%%%%%%%%%%%%%%%%%%%%%%%%%%%%%%%%%%%%
\subsection{GEUP in rainbow gravity}
%%%%%%%%%%%%%%%%%%%%%%%%%%%%%%%%%%%%%%%%%%%%%%%%%%%%%%%%%%%%%%%%%%%%%%%

Finally, by combining the momentum uncertainty (\ref{GEUP-p})
obtained from a position uncertainty of an emitted photon of order
of $\Delta x \sim r_+$ in the GEUP case with the MDR
(\ref{MDR-rainbow}) and the rainbow functions (\ref{rainbowfunc})
for a massless particle, one can also easily obtain the
GEUP-corrected black hole temperature in the rainbow gravity as
\begin{eqnarray}
 \tilde{T}^{GEUP}  &=&
 T^{GEUP}\frac{g(\omega/\omega_p)}{f(\omega/\omega_p)}\nonumber\\
                 &=&  T^{GEUP} \sqrt{1-\eta \left(\frac{\omega}{\omega_p}\right)^2}.
\end{eqnarray}
As like in the previous subsections, one can also find the energy
ratio $\omega^2/\omega^2_p$ in the square root by making use of
the momentum uncertainty (\ref{GEUP-p}) and the MDR
(\ref{MDR-rainbow}) for the massless particle, and finally obtain
the GEUP-corrected black hole temperature in the rainbow gravity
as
\begin{eqnarray}
 \tilde{T}^{GEUP} = T^{GEUP}
 \left\{1-\eta
  \left[\frac{\frac{r^2_+\omega^2_p}{2}
  \left(1-\sqrt{1-\frac{4\alpha{\cal B(\beta)}}{r^2_+\omega^2_p}}
          \right)         -\alpha{\cal B(\beta)}}
    {\alpha^2+\eta\left[\frac{r^2_+\omega^2_p}{2}
    \left(1-\sqrt{1-\frac{4\alpha{\cal B(\beta)}}{r^2_+\omega^2_p} }\right)
                        -\alpha{\cal B(\beta)}\right]}
  \right]
                       \right\}^{\frac{1}{2}}.\nonumber\\
\end{eqnarray}
Note that as $\beta\rightarrow 0$ (i.e., ${\cal
B(\beta)}\rightarrow 1$), it becomes the GUP-corrected black hole
temperature (\ref{GUP-HT}), while as $\alpha\rightarrow 0$ it goes
to the EUP-corrected black hole temperature (\ref{EUP-HT}). The
GEUP-corrected black hole temperature in the rainbow gravity is
plotted in Fig. 2(d), which shows a kind of the mixed result of
the GUP-corrected and the EUP-corrected black hole temperatures in
Fig. 2(b) and Fig. 2(c).

As like in the GUP-corrected case, the GEUP-corrected black hole
temperature in the rainbow gravity stops to evaporate at
\begin{equation}
 \tilde{T}^{GEUP}_0=\frac{\omega_p}{4\pi\sqrt{\alpha}}
                 \frac{1}{\sqrt{1-\frac{4\alpha\beta^2}
                 {\ell^2\omega^2_p}+\frac{\eta}{\alpha}}}
\end{equation}
when
\begin{equation}
 r_{+,0}=\frac{2\sqrt{\alpha}}{\omega_p}
 \frac{1}{\sqrt{1-\frac{4\alpha\beta^2}{\ell^2\omega^2_p}}}.
\end{equation}
Moreover, when $r_+\rightarrow\infty$, it approaches the value of
the EUP-corrected black hole temperature as
\begin{equation}
 \tilde{T}^{GEUP}_0=\frac{\omega_p}{4\pi\sqrt{\eta}}.
\end{equation}
It also has a local minimum at $r_+=r_m$ as shown in Fig. 2(d),
even though it is not easy to find an analytic solution. In this
respect, one can easily understand that the GEUP-corrected black
hole temperature shares the properties of the GUP-corrected and
the EUP-corrected black hole temperature: in the rainbow gravity,
the former shows that the black hole evaporation stops at some
finite $r_+$, while the latter shows that the black hole
temperature has a global minimum at some $r_+$ and approaches to a
finite value as $r_+\rightarrow 0$ and $r_+\rightarrow\infty$.

%%%%%%%%%%%%%%%%%%%%%%%%%%%%%%%%%%%%%%%%%%%%%%%%%%%%%%%%%%%%%%%%%%%%%%%
\section{Thermodynamic stability}
\setcounter{equation}{0}
\renewcommand{\theequation}{\arabic{section}.\arabic{equation}}
%%%%%%%%%%%%%%%%%%%%%%%%%%%%%%%%%%%%%%%%%%%%%%%%%%%%%%%%%%%%%%%%%%%%%%%

%%%%%%%%%%%%%%%%%%%%%%%%%%%%%%%%%%%%%%%%%%%%%%%%%%%%%%%%%%%%%%%%%%%%%%%
\subsection{Stability without rainbow gravity}
%%%%%%%%%%%%%%%%%%%%%%%%%%%%%%%%%%%%%%%%%%%%%%%%%%%%%%%%%%%%%%%%%%%%%%%

Local thermodynamic stability usually requires the positivity
condition on the specific heat. So in this subsection, let us
first study it without the rainbow gravity, and in the next
subsection, with the rainbow gravity.

To begin with, the specific heat can be simply written as
\begin{equation}
 C=\frac{dE}{dT}=\frac{dr_+}{dT}\frac{dM}{dr_+}
 =\frac{\omega^2_p}{2}\left(\frac{dT}{dr_+}\right)^{-1},
\end{equation}
where we have used the Schwarzschild black hole's mass
$M=r_+/2G=r_+\omega^2_p/2$ for simplicity. Then, the known
specific heat obtained from using the HUP is given by
\begin{equation}
 C^{HUP}=-2\pi\omega^2_pr^2_+,
\end{equation}
whose figure is drawn in Fig. 3(a) showing that in the whole range
of $r_+$ it is negative so that it is unstable thermodynamically.

On the other hand, the specific heat obtained from using the GUP
can be obtained as
\begin{equation}
 C^{GUP}=-\frac{4\pi\alpha\sqrt{1-\frac{4\alpha}{r^2_+\omega^2_p}}}
               {1-\sqrt{1-\frac{4\alpha}{r^2_+\omega^2_p}}},
\end{equation}
which is also unstable as shown in Fig. 3(b). Note that the
specific heat is only defined with $r_{+,0}\ge\Delta x_{\rm min}=
2\sqrt{\alpha}/\omega_p$.

Compared with these two, the specific heat obtained from using the
EUP given by
\begin{equation}
 C^{EUP}=-\frac{2\pi r^2_+\omega^2_p}
               {1-\frac{\beta^2}{\ell^2}r^2_+}
\end{equation}
is divided by two regions at $r_+=\ell/\beta$; when
$r_+>\ell/\beta$, it is stable, while $r_+<\ell/\beta$, it is
unstable, which is plotted in Fig. 3(c). Interestingly, the
thermodynamic behavior looks like the one of the Schwarzschild-AdS
black hole. And as $r_+\rightarrow\infty$, it goes to a constant
value of $C^{EUP}\rightarrow 2\pi\omega^2_p\ell^2/\beta^2$.

\begin{figure*}[t!]
   \centering
   \includegraphics{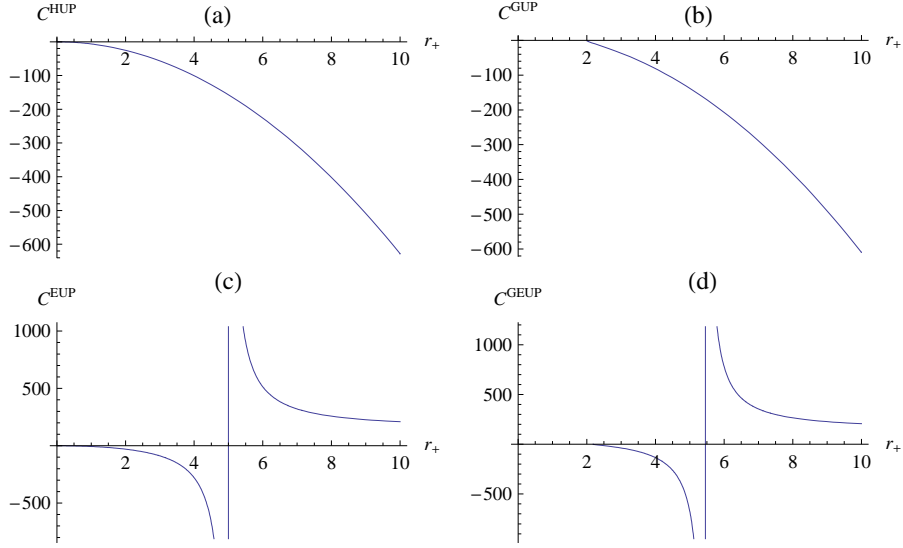}
\caption{Specific heats using (a) the HUP, (b) the GUP, (c) the
EUP, and (d) the GEUP.}
 \label{fig3}
\end{figure*}

Finally, the specific heat obtained from using the GEUP is given
by
\begin{equation}
 C^{GEUP}=-\frac{4\pi\alpha\sqrt{1-\frac{4\alpha{\cal B(\beta)}}{r^2_+\omega^2_p} }}
               {1-\frac{4\alpha\beta^2}{\ell^2\omega^2_p}
               -\sqrt{1-\frac{4\alpha{\cal B(\beta)}}{r^2_+\omega^2_p}} }.
\end{equation}
The figure 3(d) is depicted the GEUP-corrected specific heat which
is similar to the EUP-corrected specific heat $C^{EUP}$ except it
starts from $r_{+,0}=\Delta x_{\rm min}=2\sqrt{\alpha}/(\omega_p
\sqrt{1-\frac{4\alpha\beta^2}{\ell^2\omega^2_p}})$.

%%%%%%%%%%%%%%%%%%%%%%%%%%%%%%%%%%%%%%%%%%%%%%%%%%%%%%%%%%%%%%%%%%%%%%%
\subsection{Stability with rainbow gravity}
%%%%%%%%%%%%%%%%%%%%%%%%%%%%%%%%%%%%%%%%%%%%%%%%%%%%%%%%%%%%%%%%%%%%%%%

Now, let us study the thermodynamic stability by considering
specific heats with the rainbow gravity.

First, the specific heat with the HUP in the rainbow gravity is
described by
\begin{equation}
 \tilde{C}^{HUP}=-\left(\frac{2\pi(\omega^2_pr^2_++\eta)^{3/2}}{r_+\omega_p}\right),
\end{equation}
which is plotted in Fig. 4(a) showing that it is unstable and at
large $r_+$ it is reduced to the specific heat, $C^{HUP}$, of the
HUP case. Near $r_+=0$, it is a little bit different but still
remains negative, so unstable.

The GUP-corrected specific heat with the rainbow gravity can be
found as
\begin{equation}
 \tilde{C}^{GUP}=-\frac{\pi\sqrt{1-\frac{4\alpha}{r^2_+\omega^2_p}}
                \left[2\alpha(\alpha-\eta)+\eta r^2_+\omega^2_p
                \left(1-\sqrt{1-\frac{4\alpha}{r^2_+\omega^2_p}}\right)\right]^{5/2}}
               {\sqrt{2}\alpha \left[2\alpha\eta+(\alpha(\eta-\alpha)-\eta r^2_+\omega^2_p)
                   \left(1-\sqrt{1-\frac{4\alpha}{r^2_+\omega^2_p}}\right)\right]},
\end{equation}
which figure is drawn in Fig. 4(b). It shows that it starts with
$r_{+,0}=\Delta x_{\rm min}=2\sqrt{\alpha}/\omega_p$ as like in
$C^{GUP}$ and as $r_+$ becomes large, it behaves as like
$C^{GUP}$.

On the other hand, the EUP-corrected specific heat can be written
as
\begin{equation}
 \tilde{C}^{EUP}=-\frac{2\pi r^2_+\omega^2_p
                   \left[1+\frac{2\eta\beta^2}{\ell^2\omega^2_p}
                         +\frac{\eta\beta^4r^2_+}{\ell^4\omega^2_p}
                         +\frac{\eta}{r^2_+\omega^2_p}\right]^{3/2}}
               {1-\frac{\beta^2}{\ell^2}r^2_+},
\end{equation}
which has a blow-up point at $r_+=\ell/\beta$ as shown in Fig.
4(c). It is stable when $r_+>\ell/\beta$, but unstable when
$r_+<\ell/\beta$. Comparing with $C^{EUP}$, it also blows up as
$r_+\rightarrow\infty$.

\begin{figure*}[t!]
   \centering
   \includegraphics{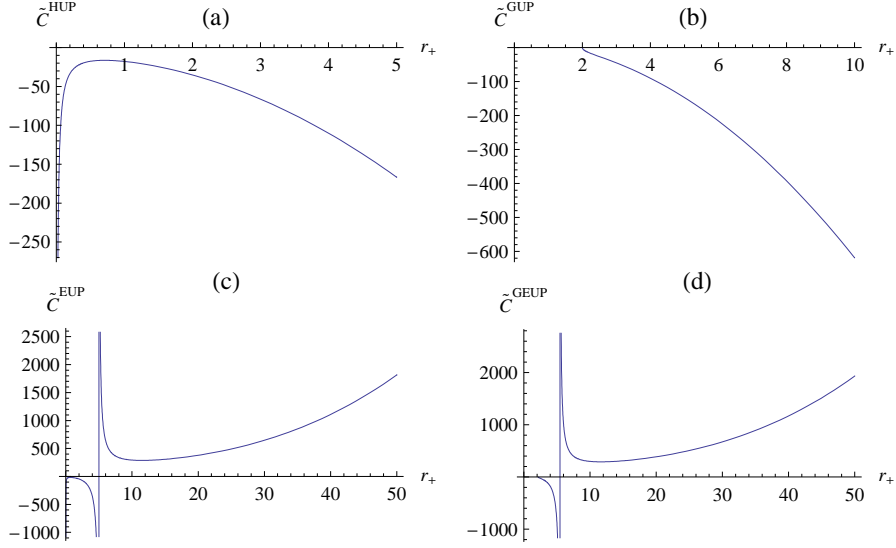}
\caption{Specific heats in the rainbow gravity using (a) the HUP,
(b) the GUP, (c) the EUP, and (d) the GEUP.}
 \label{fig4}
\end{figure*}

Finally, the GEUP-corrected specific heat is given by a little
complicated form as
\begin{equation}
 \tilde{C}^{GEUP}=-\frac{A(r_+,\omega_p,\eta,\alpha,\beta)}{B(r_+,\omega_p,\eta,\alpha,\beta)},
\end{equation}
where
\begin{eqnarray}
 A(r_+,\omega_p,\eta,\alpha,\beta)&=&\sqrt{2}\pi\left\{\frac{1-\frac{4\alpha{\cal B(\beta)}}{r^2_+\omega^2_p}}
                        {2\alpha\left(\alpha-\eta{\cal B(\beta)}\right)+\eta r^2_+\omega^2_p
         \left(1-\frac{4\alpha{\cal B(\beta)}}{r^2_+\omega^2_p}  \right)}\right\}^{\frac{1}{2}}\nonumber\\
 &\times& \left\{\frac{2\eta\alpha\beta^2r^2_+}{\ell^2}
              -\left[2\alpha(\alpha-\eta)
                                   +\eta r^2_+\omega^2_p
                                   \left(1-\frac{4\alpha{\cal B(\beta)}}{r^2_+\omega^2_p}
                                  \right)\right]\right\}^3,
                                  \nonumber\\
\end{eqnarray}
\begin{eqnarray}
B(r_+,\omega_p,\eta,\alpha,\beta)&=&\alpha^2\left\{
                        2\left[\alpha(\alpha-\eta)
                         +\eta r^2_+\omega^2_p
                        \left(1-\frac{3\alpha\beta^2}{\ell^2\omega^2_p}\right)
               \left(1-\frac{4\alpha{\cal B(\beta)}}{r^2_+\omega^2_p}\right)^{\frac{1}{2}}\right]
                         \right.\nonumber\\
                  &-&2\alpha\left[\alpha\left(1-\frac{4\alpha\beta^2}{\ell^2\omega^2_p}\right)
                    +\eta\left(3-\frac{4\alpha\beta^2}{\ell^2\omega^2_p}\right)
                    \right]\nonumber\\
                  &-&\left.2\eta\omega^2_p r^2_+\left(1-\frac{\alpha\beta^2}{\ell^2\omega^2_p}\right)
                    \left(1-\frac{4\alpha\beta^2}{\ell^2\omega^2_p}\right)
                           \right\},
\end{eqnarray}
with ${\cal B(\beta)}=1+\frac{\beta^2}{\ell^2}r^2_+$, which was
drawn in Fig. 4(d). It has similar structure with the
EUP-corrected specific heat $\tilde{C}^{EUP}$ except it starts
from $r_{+,0}=\Delta x_{\rm min}=2\sqrt{\alpha}/(\omega_p
\sqrt{1-\frac{4\alpha\beta^2}{\ell^2\omega^2_p}})$ as before.

%%%%%%%%%%%%%%%%%%%%%%%%%%%%%%%%%%%%%%%%%%%%%%%%%%%%%%%%%%%%%%%%%%%%%%%
\section{Discussion}
\setcounter{equation}{0}
\renewcommand{\theequation}{\arabic{section}.\arabic{equation}}
%%%%%%%%%%%%%%%%%%%%%%%%%%%%%%%%%%%%%%%%%%%%%%%%%%%%%%%%%%%%%%%%%%%%%%%

In this paper, according to the method of Adler-Chen-Santiago, we
have studied modified temperatures of black holes in the rainbow
gravity. By fully considering both the MDR and the three different
types of the GUPs including the extended uncertainty principle
(EUP) and the generalized extended uncertainty principle (GEUP),
we have eliminated the energy dependence of a particle in the
modified black hole temperature in the rainbow gravity. As a
result, we have shown that the rainbow gravity makes the black
hole temperature finite at $r_+=0$ in the HUP-corrected and the
EUP-corrected case, while in the GUP-corrected and the
GEUP-corrected case the black hole temperatures are also finite
not at $r_{+,0}=0$, but at $r_{+,0}=2\sqrt{\alpha}/\omega_p$ and
at $r_{+,0}=2\sqrt{\alpha}/(\omega_p
\sqrt{1-\frac{4\alpha\beta^2}{\ell^2\omega^2_p}})$ where they stop
the black hole evaporation.

We have also investigated their corresponding thermodynamic
stabilities of the various GUPs-corrected modified Schwarzschild
black holes in the rainbow gravity. The HUP-corrected and the
GUP-corrected black holes are globally unstable. On the other
hand, the EUP-corrected and the GEUP-corrected black holes are
stable in a certain region of satisfying with $r_+>r_m$, while
they are unstable when $r_+<r_m$.

Through a further investigation, it will be interesting to analyze
thermodynamic phase transitions in the rainbow gravity by fully
implementing the MDR and the various GUPs.

\section*{Acknowledgments}

Y.-W. Kim was supported by Basic Science Research Program through
the National Research Foundation of Korea (NRF) funded by the
Ministry of Education, Science and Technology
(NRF-2017R1A2B4011702). Y.-J. Park was supported by the Sogang
University Research Grant of 2016 (201610102.01.).

%%%%%%%%%%%%%%%%%%%%%%%%%%%%%%%%%%%%%%%%%%%%%%%%%%%%%%%%%%%%%
%%%%%%%%%%%%%%%             References       %%%%%%%%%%%%%%%%
%%%%%%%%%%%%%%%%%%%%%%%%%%%%%%%%%%%%%%%%%%%%%%%%%%%%%%%%%%%%%


\begin{thebibliography}{99}

%1
\bibitem{Hawking:1974sw}
  S.~W.~Hawking,
  ``Particle Creation by Black Holes,''
  Commun.\ Math.\ Phys.\  {\bf 43}, 199 (1975)
  [Commun.\ Math.\ Phys.\  {\bf 46}, 206 (1976)].
  %%CITATION = CMPHA,43,199;%%
  %5605 citations counted in INSPIRE as of 03 sept. 2015


%%%%%%%%%%%%%%%%%%%%%%%%%%%%%%%%%%%%%%%%%%%%%
% GUP thought experiments
%%%%%%%%%%%%%%%%%%%%%%%%%%%%%%%%%%%%%%%%%%%%%

%2
%\cite{Maggiore:1993rv}
\bibitem{Maggiore:1993rv}
  M.~Maggiore,
  ``A Generalized uncertainty principle in quantum gravity,''
  Phys.\ Lett.\ B {\bf 304}, 65 (1993)
 % doi:10.1016/0370-2693(93)91401-8
  [hep-th/9301067].
  %%CITATION = doi:10.1016/0370-2693(93)91401-8;%%
  %484 citations counted in INSPIRE as of 11 Jan 2017

%3
%\cite{Scardigli:1999jh}
\bibitem{Scardigli:1999jh}
  F.~Scardigli,
  ``Generalized uncertainty principle in quantum gravity from micro - black hole Gedanken experiment,''
  Phys.\ Lett.\ B {\bf 452}, 39 (1999)
  %doi:10.1016/S0370-2693(99)00167-7
  [hep-th/9904025].
  %%CITATION = doi:10.1016/S0370-2693(99)00167-7;%%
  %342 citations counted in INSPIRE as of 11 Jan 2017


%%%%%%%%%%%%%%%%%%%%%%%%%%%%%%%%%%%%%%%%%%%%%
% GUP
%%%%%%%%%%%%%%%%%%%%%%%%%%%%%%%%%%%%%%%%%%%%%

%4
%\cite{Amati:1988tn}
\bibitem{Amati:1988tn}
  D.~Amati, M.~Ciafaloni and G.~Veneziano,
  ``Can Space-Time Be Probed Below the String Size?,''
  Phys.\ Lett.\ B {\bf 216}, 41 (1989).
  %doi:10.1016/0370-2693(89)91366-X
  %%CITATION = doi:10.1016/0370-2693(89)91366-X;%%
  %867 citations counted in INSPIRE as of 07 Sep 2017

%5
%\cite{Konishi:1989wk}
\bibitem{Konishi:1989wk}
  K.~Konishi, G.~Paffuti and P.~Provero,
  ``Minimum Physical Length and the Generalized Uncertainty Principle in String Theory,''
  Phys.\ Lett.\ B {\bf 234}, 276 (1990).
 % doi:10.1016/0370-2693(90)91927-4
  %%CITATION = doi:10.1016/0370-2693(90)91927-4;%%
  %382 citations counted in INSPIRE as of 07 Sep 2017

%6
%\cite{Kato:1990bd}
\bibitem{Kato:1990bd}
  M.~Kato,
  ``Particle Theories With Minimum Observable Length and Open String Theory,''
  Phys.\ Lett.\ B {\bf 245}, 43 (1990).
  %doi:10.1016/0370-2693(90)90162-Y
  %%CITATION = doi:10.1016/0370-2693(90)90162-Y;%%
  %88 citations counted in INSPIRE as of 11 Sep 2017

%7
%\cite{Kempf:1993bq}
\bibitem{Kempf:1993bq}
  A.~Kempf,
  ``Uncertainty relation in quantum mechanics with quantum group symmetry,''
  J.\ Math.\ Phys.\  {\bf 35}, 4483 (1994)
  %doi:10.1063/1.530798
  [hep-th/9311147].
  %%CITATION = doi:10.1063/1.530798;%%
  %239 citations counted in INSPIRE as of 07 Sep 2017

%8
%\cite{Garay:1994en}
\bibitem{Garay:1994en}
  L.~J.~Garay,
  ``Quantum gravity and minimum length,''
  Int.\ J.\ Mod.\ Phys.\ A {\bf 10}, 145 (1995)
 % doi:10.1142/S0217751X95000085
  [gr-qc/9403008].
  %%CITATION = doi:10.1142/S0217751X95000085;%%
  %841 citations counted in INSPIRE as of 11 Jan 2017

%9
%\cite{Kempf:1996nk}
\bibitem{Kempf:1996nk}
  A.~Kempf and G.~Mangano,
  ``Minimal length uncertainty relation and ultraviolet regularization,''
  Phys.\ Rev.\ D {\bf 55}, 7909 (1997)
%  doi:10.1103/PhysRevD.55.7909
  [hep-th/9612084].
  %%CITATION = doi:10.1103/PhysRevD.55.7909;%%
  %286 citations counted in INSPIRE as of 06 Sep 2017

%10
%\cite{KalyanaRama:2001xd}
\bibitem{KalyanaRama:2001xd}
  S.~Kalyana Rama,
  ``Some consequences of the generalized uncertainty principle: Statistical mechanical, cosmological, and varying speed of light,''
  Phys.\ Lett.\ B {\bf 519}, 103 (2001)
  %doi:10.1016/S0370-2693(01)01091-7
  [hep-th/0107255].
  %%CITATION = doi:10.1016/S0370-2693(01)01091-7;%%
  %91 citations counted in INSPIRE as of 06 Sep 2017

%11
%\cite{Chang:2001bm}
\bibitem{Chang:2001bm}
  L.~N.~Chang, D.~Minic, N.~Okamura and T.~Takeuchi,
  ``The Effect of the minimal length uncertainty relation on the density of states and the cosmological constant problem,''
  Phys.\ Rev.\ D {\bf 65}, 125028 (2002)
 % doi:10.1103/PhysRevD.65.125028
  [hep-th/0201017].
  %%CITATION = doi:10.1103/PhysRevD.65.125028;%%
  %256 citations counted in INSPIRE as of 06 Sep 2017

%12
%\cite{Medved:2004yu}
\bibitem{Medved:2004yu}
  A.~J.~M.~Medved and E.~C.~Vagenas,
  ``When conceptual worlds collide: The GUP and the BH entropy,''
  Phys.\ Rev.\ D {\bf 70}, 124021 (2004)
  %doi:10.1103/PhysRevD.70.124021
  [hep-th/0411022].
  %%CITATION = doi:10.1103/PhysRevD.70.124021;%%
  %208 citations counted in INSPIRE as of 06 Sep 2017


%13
%\cite{Setare:2005sj}
\bibitem{Setare:2005sj}
  M.~R.~Setare,
  ``The Generalized uncertainty principle and corrections to the Cardy-Verlinde formula in SAdS(5) black holes,''
  Int.\ J.\ Mod.\ Phys.\ A {\bf 21}, 1325 (2006)
  %doi:10.1142/S0217751X06025304
  [hep-th/0504179].
  %%CITATION = doi:10.1142/S0217751X06025304;%%
  %51 citations counted in INSPIRE as of 06 Sep 2017

%14
%\cite{Nouicer:2005dp}
\bibitem{Nouicer:2005dp}
  K.~Nouicer,
  ``Casimir effect in the presence of minimal lengths,''
  J.\ Phys.\ A {\bf 38}, 10027 (2005)
  % doi:10.1088/0305-4470/38/46/009
  [hep-th/0512027].
  %%CITATION = doi:10.1088/0305-4470/38/46/009;%%
  %35 citations counted in INSPIRE as of 06 Sep 2017

%15
%\cite{Kim:2006rx}
\bibitem{Kim:2006rx}
  W.~Kim, Y.~W.~Kim and Y.~J.~Park,
  ``Entropy of the Randall-Sundrum brane world with the generalized uncertainty principle,''
  Phys.\ Rev.\ D {\bf 74}, 104001 (2006)
  %doi:10.1103/PhysRevD.74.104001
  [gr-qc/0605084].
  %%CITATION = doi:10.1103/PhysRevD.74.104001;%%
  %64 citations counted in INSPIRE as of 06 Sep 2017

%16
%\cite{Myung:2006qr}
\bibitem{Myung:2006qr}
  Y.~S.~Myung, Y.~W.~Kim and Y.~J.~Park,
  ``Black hole thermodynamics with generalized uncertainty principle,''
  Phys.\ Lett.\ B {\bf 645}, 393 (2007)
  %doi:10.1016/j.physletb.2006.12.062
  [gr-qc/0609031].
  %%CITATION = doi:10.1016/j.physletb.2006.12.062;%%
  %69 citations counted in INSPIRE as of 06 Sep 2017

%17
%\cite{Kim:2007if}
\bibitem{Kim:2007if}
  Y.~W.~Kim and Y.~J.~Park,
  ``Entropy of the Schwarzschild black hole to all orders in the Planck length,''
  Phys.\ Lett.\ B {\bf 655}, 172 (2007)
  %doi:10.1016/j.physletb.2007.08.065
  [arXiv:0707.2128 [gr-qc]].
  %%CITATION = doi:10.1016/j.physletb.2007.08.065;%%
  %60 citations counted in INSPIRE as of 06 Sep 2017


%18
%\cite{Nozari:2010qy}
\bibitem{Nozari:2010qy}
  K.~Nozari and P.~Pedram,
  ``Minimal Length and Bouncing Particle Spectrum,''
  EPL {\bf 92}, 50013 (2010)
  %doi:10.1209/0295-5075/92/50013
  [arXiv:1011.5673 [hep-th]].
  %%CITATION = doi:10.1209/0295-5075/92/50013;%%
  %43 citations counted in INSPIRE as of 06 Sep 2017

%19
%\cite{Ali:2011ap}
\bibitem{Ali:2011ap}
  A.~F.~Ali,
  ``Minimal Length in Quantum Gravity, Equivalence Principle and Holographic Entropy Bound,''
  Class.\ Quant.\ Grav.\  {\bf 28}, 065013 (2011)
 % doi:10.1088/0264-9381/28/6/065013
  [arXiv:1101.4181 [hep-th]].
  %%CITATION = doi:10.1088/0264-9381/28/6/065013;%%
  %55 citations counted in INSPIRE as of 06 Sep 2017

%20
%\cite{Pedram:2011gw}
\bibitem{Pedram:2011gw}
  P.~Pedram,
  ``A Higher Order GUP with Minimal Length Uncertainty and Maximal Momentum,''
  Phys.\ Lett.\ B {\bf 714}, 317 (2012)
  %doi:10.1016/j.physletb.2012.07.005
  [arXiv:1110.2999 [hep-th]].
  %%CITATION = doi:10.1016/j.physletb.2012.07.005;%%
  %45 citations counted in INSPIRE as of 06 Sep 2017

%21
%\cite{Sprenger:2012uc}
\bibitem{Sprenger:2012uc}
  M.~Sprenger, P.~Nicolini and M.~Bleicher,
  ``Physics on Smallest Scales - An Introduction to Minimal Length Phenomenology,''
  Eur.\ J.\ Phys.\  {\bf 33}, 853 (2012)
  %doi:10.1088/0143-0807/33/4/853
  [arXiv:1202.1500 [physics.ed-ph]].
  %%CITATION = doi:10.1088/0143-0807/33/4/853;%%
  %51 citations counted in INSPIRE as of 06 Sep 2017

%22
%\cite{Hossenfelder:2012jw}
\bibitem{Hossenfelder:2012jw}
  S.~Hossenfelder,
  ``Minimal Length Scale Scenarios for Quantum Gravity,''
  Living Rev.\ Rel.\  {\bf 16}, 2 (2013)
 % doi:10.12942/lrr-2013-2
  [arXiv:1203.6191 [gr-qc]].
  %%CITATION = doi:10.12942/lrr-2013-2;%%
  %182 citations counted in INSPIRE as of 11 Jan 2017


%23
%\cite{Dehghani:2016yjm}
\bibitem{Dehghani:2016yjm}
  M.~Dehghani,
  ``The Cardy-Verlinde formula in the presence of GUP and MDR,''
  Astrophys.\ Space Sci.\  {\bf 361}, 148 (2016).
  %doi:10.1007/s10509-016-2727-y
  %%CITATION = doi:10.1007/s10509-016-2727-y;%%

%24
%\cite{Zhou:2016cwy}
\bibitem{Zhou:2016cwy}
  S.~Zhou and G.~R.~Chen,
  ``Corrected black hole thermodynamics in Damour-Ruffini¡¯s method with generalized uncertainty principle,''
  Int.\ J.\ Mod.\ Phys.\ D {\bf 26}, 1750062 (2016).
  %doi:10.1142/S0218271817500626
  %%CITATION = doi:10.1142/S0218271817500626;%%

%25
%\cite{Bernardo:2016dcy}
\bibitem{Bernardo:2016dcy}
  R.~C.~S.~Bernardo and J.~P.~H.~Esguerra,
  ``Quantum scattering in one-dimensional systems satisfying the minimal length uncertainty relation,''
  Annals Phys.\  {\bf 375}, 444 (2016).
 % doi:10.1016/j.aop.2016.10.022
  %%CITATION = doi:10.1016/j.aop.2016.10.022;%%

%26
%\cite{Feng:2016tyt}
\bibitem{Feng:2016tyt}
  Z.~W.~Feng, S.~Z.~Yang, H.~L.~Li and X.~T.~Zu,
  ``Constraining the generalized uncertainty principle with the gravitational wave event GW150914,''
  Phys.\ Lett.\ B {\bf 768}, 81 (2017)
 % doi:10.1016/j.physletb.2017.02.043
  [arXiv:1610.08549 [hep-ph]].
  %%CITATION = doi:10.1016/j.physletb.2017.02.043;%%
  %3 citations counted in INSPIRE as of 07 Sep 2017

%27
%\cite{Tao:2017mpe}
\bibitem{Tao:2017mpe}
  J.~Tao, P.~Wang and H.~Yang,
  ``Black hole radiation with modified dispersion relation in tunneling paradigm: Static frame,''
  Nucl.\ Phys.\ B {\bf 922}, 346 (2017).
  %doi:10.1016/j.nuclphysb.2017.06.022
  %%CITATION = doi:10.1016/j.nuclphysb.2017.06.022;%%


%%%%%%%%%%%%%%%%%%%%%%%%%%%%%%%%%%%%%%%%%%%%%%%%%%%%%%%%%%%%%%%%%%%%%%%%%%%%%%
%  GEUP
%%%%%%%%%%%%%%%%%%%%%%%%%%%%%%%%%%%%%%%%%%%%%%%%%%%%%%%%%%%%%%%%%%%%%%%%%%%%%%

%28
%\cite{Kempf:1994su}
\bibitem{Kempf:1994su}
  A.~Kempf, G.~Mangano and R.~B.~Mann,
  ``Hilbert space representation of the minimal length uncertainty relation,''
  Phys.\ Rev.\ D {\bf 52}, 1108 (1995)
%  doi:10.1103/PhysRevD.52.1108
  [hep-th/9412167].
  %%CITATION = doi:10.1103/PhysRevD.52.1108;%%
  %887 citations counted in INSPIRE as of 06 Sep 2017

%29
%\cite{Tawfik:2014zca}
\bibitem{Tawfik:2014zca}
  A.~N.~Tawfik and A.~M.~Diab,
  ``Generalized Uncertainty Principle: Approaches and Applications,''
  Int.\ J.\ Mod.\ Phys.\ D {\bf 23}, 1430025 (2014)
 % doi:10.1142/S0218271814300250
  [arXiv:1410.0206 [gr-qc]].
  %%CITATION = doi:10.1142/S0218271814300250;%%
  %38 citations counted in INSPIRE as of 06 Sep 2017

%30
%\cite{Anacleto:2015rlz}
\bibitem{Anacleto:2015rlz}
  M.~A.~Anacleto, D.~Bazeia, F.~A.~Brito and J.~C.~Mota-Silva,
  ``Quantum-corrected two-dimensional Horava-Lifshitz black hole entropy,''
  Adv.\ High Energy Phys.\  {\bf 2016}, 8465759 (2016)
  %doi:10.1155/2016/8465759
  [arXiv:1512.07886 [hep-th]].
  %%CITATION = doi:10.1155/2016/8465759;%%
  %2 citations counted in INSPIRE as of 07 Sep 2017

%31
%\cite{Bolen:2004sq}
\bibitem{Bolen:2004sq}
  B.~Bolen and M.~Cavaglia,
  ``(Anti-)de Sitter black hole thermodynamics and the generalized uncertainty principle,''
  Gen.\ Rel.\ Grav.\  {\bf 37}, 1255 (2005)
  %doi:10.1007/s10714-005-0108-x
  [gr-qc/0411086].
  %%CITATION = doi:10.1007/s10714-005-0108-x;%%
  %63 citations counted in INSPIRE as of 06 Sep 2017

%32
%\cite{Bambi:2007ty}
\bibitem{Bambi:2007ty}
  C.~Bambi and F.~R.~Urban,
  ``Natural extension of the Generalised Uncertainty Principle,''
  Class.\ Quant.\ Grav.\  {\bf 25}, 095006 (2008)
  %doi:10.1088/0264-9381/25/9/095006
  [arXiv:0709.1965 [gr-qc]].
  %%CITATION = doi:10.1088/0264-9381/25/9/095006;%%
  %111 citations counted in INSPIRE as of 12 Sep 2017

%33
%\cite{Park:2007az}
\bibitem{Park:2007az}
  M.~I.~Park,
  ``The Generalized Uncertainty Principle in (A)dS Space and the Modification of Hawking Temperature from the Minimal Length,''
  Phys.\ Lett.\ B {\bf 659}, 698 (2008)
  %doi:10.1016/j.physletb.2007.11.090
  [arXiv:0709.2307 [hep-th]].
  %%CITATION = doi:10.1016/j.physletb.2007.11.090;%%
  %76 citations counted in INSPIRE as of 06 Sep 2017

%%%%%%%%%%%%%%%%%%%%%%%%%%%%%%%%%%%%%%%%%%%%%%%%%%%%%%%%%%%%%%%%%%%%%%%
% ACS
%%%%%%%%%%%%%%%%%%%%%%%%%%%%%%%%%%%%%%%%%%%%%%%%%%%%%%%%%%%%%%%%%%%%%%%%

%34
%\cite{Adler:2001vs}
\bibitem{Adler:2001vs}
  R.~J.~Adler, P.~Chen and D.~I.~Santiago,
  ``The Generalized uncertainty principle and black hole remnants,
  Gen.\ Rel.\ Grav.\  {\bf 33}, 2101 (2001)
  [gr-qc/0106080].
  %%CITATION = doi:10.1023/A:1015281430411;%%
  %322 citations counted in INSPIRE as of 08 Nov 2016

%%%%%%%%%%%%%%%%%%%%%%%%%%%%%%%%%%%%%%%%%%%%%%%%%%%%%%%%%%%%%%%%%%%%%%%%
% Double special relativity
%%%%%%%%%%%%%%%%%%%%%%%%%%%%%%%%%%%%%%%%%%%%%%%%%%%%%%%%%%%%%%%%%%%%%%%%

%35
%\cite{AmelinoCamelia:2000ge}
\bibitem{AmelinoCamelia:2000ge}
  G.~Amelino-Camelia,
  ``Testable scenario for relativity with minimum length,''
  Phys.\ Lett.\ B {\bf 510}, 255 (2001)
%  doi:10.1016/S0370-2693(01)00506-8
  [hep-th/0012238].
  %%CITATION = doi:10.1016/S0370-2693(01)00506-8;%%
  %575 citations counted in INSPIRE as of 21 Aug 2017

%36
%\cite{AmelinoCamelia:2000mn}
\bibitem{AmelinoCamelia:2000mn}
  G.~Amelino-Camelia,
  ``Relativity in space-times with short distance structure governed by an observer independent (Planckian) length scale,''
  Int.\ J.\ Mod.\ Phys.\ D {\bf 11}, 35 (2002)
%  doi:10.1142/S0218271802001330
  [gr-qc/0012051].
  %%CITATION = doi:10.1142/S0218271802001330;%%
  %847 citations counted in INSPIRE as of 21 Aug 2017

%37
%\cite{AmelinoCamelia:2002wr}
\bibitem{AmelinoCamelia:2002wr}
  G.~Amelino-Camelia,
  ``Doubly special relativity,''
  Nature {\bf 418}, 34 (2002)
%  doi:10.1038/418034a
  [gr-qc/0207049].
  %%CITATION = doi:10.1038/418034a;%%
  %273 citations counted in INSPIRE as of 06 Sep 2017

%38
%\cite{KowalskiGlikman:2004qa}
\bibitem{KowalskiGlikman:2004qa}
  J.~Kowalski-Glikman,
  ``Introduction to doubly special relativity,''
  Lect.\ Notes Phys.\  {\bf 669}, 131 (2005)
  %doi:10.1007/11377306_5
  [hep-th/0405273].
  %%CITATION = doi:10.1007/11377306_5;%%
  %163 citations counted in INSPIRE as of 06 Sep 2017


%%%%%%%%%%%%%%%%%%%%%%%%%%%%%%%%%%%%%%%%%%%%%%%%%%%%
% Rainbow
%%%%%%%%%%%%%%%%%%%%%%%%%%%%%%%%%%%%%%%%%%%%%%%%%%%%

%39
%\cite{Magueijo:2002xx}
\bibitem{Magueijo:2002xx}
  J.~Magueijo and L.~Smolin,
   ``Gravity's rainbow,''
  Class.\ Quant.\ Grav.\  {\bf 21}, 1725 (2004)
  [gr-qc/0305055].
  %%CITATION = GR-QC/0305055;%%
  %193 citations counted in INSPIRE as of 03 sept. 2015

%40
%\cite{Liberati:2004ju}
\bibitem{Liberati:2004ju}
  S.~Liberati, S.~Sonego and M.~Visser,
  ``Interpreting doubly special relativity as a modified theory of measurement,''
  Phys.\ Rev.\ D {\bf 71}, 045001 (2005)
  [gr-qc/0410113].
  %%CITATION = GR-QC/0410113;%%
  %42 citations counted in INSPIRE as of 18 Aug 2015

%41
\bibitem{Galan:2004st}
  P.~Galan and G.~A.~Mena Marugan,
  ``Quantum time uncertainty in a gravity's rainbow formalism,''
  Phys.\ Rev.\ D {\bf 70}, 124003 (2004)
  [gr-qc/0411089].
  %%CITATION = GR-QC/0411089;%%
  %27 citations counted in INSPIRE as of 11 May 2014

%42
%\cite{Galan:2005ju}
\bibitem{Galan:2005ju}
  P.~Galan and G.~A.~Mena Marugan,
 ``Length uncertainty in a gravity's rainbow formalism,''
  Phys.\ Rev.\ D {\bf 72}, 044019 (2005)
  [gr-qc/0507098].
  %%CITATION = GR-QC/0507098;%%
  %18 citations counted in INSPIRE as of 11 May 2014

%43
%\cite{Hackett:2005mb}
\bibitem{Hackett:2005mb}
  J.~Hackett,
  ``Asymptotic flatness in rainbow gravity,''
  Class.\ Quant.\ Grav.\  {\bf 23}, 3833 (2006)
  [gr-qc/0509103].
  %%CITATION = GR-QC/0509103;%%
  %19 citations counted in INSPIRE as of 11 May 2014

%44
%\cite{Ling:2006az}
\bibitem{Ling:2006az}
  Y.~Ling,
  ``Rainbow universe,''
  JCAP {\bf 0708}, 017 (2007)
  [gr-qc/0609129].
  %%CITATION = GR-QC/0609129;%%
  %14 citations counted in INSPIRE as of 11 May 2014

%45
%\cite{Ling:2006ba}
\bibitem{Ling:2006ba}
  Y.~Ling, S.~He and H.~-b.~Zhang,
  ``The Kinematics of particles moving in rainbow spacetime,''
  Mod.\ Phys.\ Lett.\ A {\bf 22}, 2931 (2007)
  [gr-qc/0609130].
  %%CITATION = GR-QC/0609130;%%
  %7 citations counted in INSPIRE as of 11 May 2014

%46
%\cite{Girelli:2006fw}
\bibitem{Girelli:2006fw}
  F.~Girelli, S.~Liberati and L.~Sindoni,
  ``Planck-scale modified dispersion relations and Finsler geometry,''
  Phys.\ Rev.\ D {\bf 75}, 064015 (2007)
  [gr-qc/0611024].
  %%CITATION = GR-QC/0611024;%%
  %101 citations counted in INSPIRE as of 11 May 2014

%47
%\cite{Li:2008gs}
\bibitem{Li:2008gs}
  H.~Li, Y.~Ling and X.~Han,
  ``Modified (A)dS Schwarzschild black holes in Rainbow spacetime,''
  Class.\ Quant.\ Grav.\  {\bf 26}, 065004 (2009)
  %doi:10.1088/0264-9381/26/6/065004
  [arXiv:0809.4819 [gr-qc]].
  %%CITATION = doi:10.1088/0264-9381/26/6/065004;%%
  %37 citations counted in INSPIRE as of 06 Sep 2017

%48
%\cite{Ling:2008sy}
\bibitem{Ling:2008sy}
  Y.~Ling and Q.~Wu,
  ``The Big Bounce in Rainbow Universe,''
  Phys.\ Lett.\ B {\bf 687}, 103 (2010)
  [arXiv:0811.2615 [gr-qc]].
  %%CITATION = ARXIV:0811.2615;%%
  %17 citations counted in INSPIRE as of 03 sept. 2015

%49
%\cite{Garattini:2011kp}
\bibitem{Garattini:2011kp}
  R.~Garattini and G.~Mandanici,
  ``Modified Dispersion Relations lead to a finite Zero Point Gravitational Energy,''
  Phys.\ Rev.\ D {\bf 83}, 084021 (2011)
  [arXiv:1102.3803 [gr-qc]].
  %%CITATION = ARXIV:1102.3803;%%
  %33 citations counted in INSPIRE as of 03 sept. 2015



%50
%\cite{Garattini:2011fs}
\bibitem{Garattini:2011fs}
  R.~Garattini and F.~S.~N.~Lobo,
  ``Self-sustained wormholes in modified dispersion relations,''
  Phys.\ Rev.\ D {\bf 85}, 024043 (2012)
  [arXiv:1111.5729 [gr-qc]].
  %%CITATION = ARXIV:1111.5729;%%
  %23 citations counted in INSPIRE as of 03 sept. 2015

%51
%\cite{Garattini:2012ec}
\bibitem{Garattini:2012ec}
  R.~Garattini,
  ``Distorting General Relativity: Gravity's Rainbow and f(R) theories at work,''
  JCAP {\bf 1306}, 017 (2013)
 % doi:10.1088/1475-7516/2013/06/017
  [arXiv:1210.7760 [gr-qc]].
  %%CITATION = doi:10.1088/1475-7516/2013/06/017;%%
  %14 citations counted in INSPIRE as of 18 May 2016

%52
%\cite{Garattini:2012ca}
\bibitem{Garattini:2012ca}
  R.~Garattini and M.~Sakellariadou,
  ``Does gravity's rainbow induce inflation without an inflaton?,''
  Phys.\ Rev.\ D {\bf 90}, 043521 (2014)
  % doi:10.1103/PhysRevD.90.043521
  [arXiv:1212.4987 [gr-qc]].
  %%CITATION = doi:10.1103/PhysRevD.90.043521;%%
  %13 citations counted in INSPIRE as of 18 May 2016

%53
%\cite{Majumder:2013mza}
\bibitem{Majumder:2013mza}
  B.~Majumder,
  ``Singularity Free Rainbow Universe,''
  Int.\ J.\ Mod.\ Phys.\ D {\bf 22}, 1342021 (2013)
 % doi:10.1142/S0218271813420212
  [arXiv:1305.3709 [gr-qc]].
  %%CITATION = doi:10.1142/S0218271813420212;%%
  %7 citations counted in INSPIRE as of 18 May 2016

%54
%\cite{Amelino-Camelia:2013wha}
\bibitem{Amelino-Camelia:2013wha}
  G.~Amelino-Camelia, M.~Arzano, G.~Gubitosi and J.~Magueijo,
  ``Rainbow gravity and scale-invariant fluctuations,''
  Phys.\ Rev.\ D {\bf 88}, 041303 (2013)
  [arXiv:1307.0745 [gr-qc]].
  %%CITATION = ARXIV:1307.0745;%%
  %17 citations counted in INSPIRE as of 03 sept. 2015

%55
%\cite{Awad:2013nxa}
\bibitem{Awad:2013nxa}
  A.~Awad, A.~F.~Ali and B.~Majumder,
  ``Nonsingular Rainbow Universes,''
  JCAP {\bf 1310}, 052 (2013)
  [arXiv:1308.4343 [gr-qc]].
  %%CITATION = ARXIV:1308.4343;%%
  %20 citations counted in INSPIRE as of 03 sept. 2015

%56
%\cite{Barrow:2013gia}
\bibitem{Barrow:2013gia}
  J.~D.~Barrow and J.~Magueijo,
  ``Intermediate inflation from rainbow gravity,''
  Phys.\ Rev.\ D {\bf 88}, 103525 (2013)
  [arXiv:1310.2072 [astro-ph.CO]].
  %%CITATION = ARXIV:1310.2072;%%
  %17 citations counted in INSPIRE as of 03 sept. 2015


%57
%\cite{Ali:2014zea}
\bibitem{Ali:2014zea}
  A.~F.~Ali, M.~Faizal and M.~M.~Khalil,
  ``Remnant for all Black Objects due to Gravity's Rainbow,''
  Nucl.\ Phys.\ B {\bf 894}, 341 (2015)
  %doi:10.1016/j.nuclphysb.2015.03.014
  [arXiv:1410.5706 [hep-th]].
  %%CITATION = doi:10.1016/j.nuclphysb.2015.03.014;%%
  %53 citations counted in INSPIRE as of 06 Sep 2017

%58
%\cite{Gim:2015zra}
\bibitem{Gim:2015zra}
  Y.~Gim and W.~Kim,
  ``Black Hole Complementarity in Gravity's Rainbow,''
  JCAP {\bf 1505}, 002 (2015)
  %doi:10.1088/1475-7516/2015/05/002
  [arXiv:1501.04702 [gr-qc]].
  %%CITATION = doi:10.1088/1475-7516/2015/05/002;%%
  %21 citations counted in INSPIRE as of 06 Sep 2017

%59
%\cite{Santos:2015sva}
\bibitem{Santos:2015sva}
  G.~Santos, G.~Gubitosi and G.~Amelino-Camelia,
  ``On the initial singularity problem in rainbow cosmology,''
  JCAP {\bf 1508}, 005 (2015)
 % doi:10.1088/1475-7516/2015/08/005
  [arXiv:1502.02833 [gr-qc]].
  %%CITATION = doi:10.1088/1475-7516/2015/08/005;%%
  %6 citations counted in INSPIRE as of 18 May 2016

%60
%\cite{Hendi:2015cra}
\bibitem{Hendi:2015cra}
  S.~H.~Hendi, M.~Faizal, B.~E.~Panah and S.~Panahiyan,
  ``Charged dilatonic black holes in gravity¡¯s rainbow,''
  Eur.\ Phys.\ J.\ C {\bf 76}, 296 (2016)
  %doi:10.1140/epjc/s10052-016-4119-4
  [arXiv:1508.00234 [hep-th]].
  %%CITATION = doi:10.1140/epjc/s10052-016-4119-4;%%
  %29 citations counted in INSPIRE as of 06 Sep 2017

%61
%\cite{Kim:2015wwa}
\bibitem{Kim:2015wwa}
  Y.~W.~Kim and Y.~J.~Park,
  ``Local free-fall Temperature of modified Schwarzschild black hole in rainbow spacetime,''
  Mod.\ Phys.\ Lett.\ A {\bf 31}, 1650106 (2016)
  %doi:10.1142/S0217732316501066
  [arXiv:1508.07439 [gr-qc]].
  %%CITATION = doi:10.1142/S0217732316501066;%%
  %4 citations counted in INSPIRE as of 06 Sep 2017

%62
%\cite{Gim:2015yxa}
\bibitem{Gim:2015yxa}
  Y.~Gim and W.~Kim,
  ``Hawking, fiducial, and free-fall temperature of black hole on gravity¡¯s rainbow,''
  Eur.\ Phys.\ J.\ C {\bf 76}, 166 (2016)
  %doi:10.1140/epjc/s10052-016-4025-9
  [arXiv:1509.06846 [gr-qc]].
  %%CITATION = doi:10.1140/epjc/s10052-016-4025-9;%%
  %12 citations counted in INSPIRE as of 06 Sep 2017

%63
%\cite{Carvalho:2015omv}
\bibitem{Carvalho:2015omv}
  G.~G.~Carvalho, I.~P.~Lobo and E.~Bittencourt,
  ``Extended disformal approach in the scenario of Rainbow Gravity,''
  Phys.\ Rev.\ D {\bf 93}, 044005 (2016)
 % doi:10.1103/PhysRevD.93.044005
  [arXiv:1511.00495 [gr-qc]].
  %%CITATION = doi:10.1103/PhysRevD.93.044005;%%
  %3 citations counted in INSPIRE as of 18 May 2016

%64
\bibitem{Hendi:1512}
  S.~H.~Hendi, S.~Panahiyan, B.~E.~Panah and M.~Momennia,
  ``Thermodynamics instability of nonlinearly charged black holes in gravity¡¯s rainbow,''
  Eur.\ Phys.\ J.\ C {\bf 76}, 150 (2016)
  [arXiv:1512.05192 [physics]].

%65
%\cite{Hendi:2016vux}
\bibitem{Hendi:2016vux}
  S.~H.~Hendi, B.~Eslam Panah and S.~Panahiyan,
  ``Topological charged black holes in massive gravity's rainbow and their thermodynamical analysis through various approaches,''
  Phys.\ Lett.\ B {\bf 769}, 191 (2017)
  %doi:10.1016/j.physletb.2017.03.051
  [arXiv:1602.01832 [gr-qc]].
  %%CITATION = doi:10.1016/j.physletb.2017.03.051;%%
  %17 citations counted in INSPIRE as of 06 Sep 2017

%66
%\cite{Ashour:2016cay}
\bibitem{Ashour:2016cay}
  A.~Ashour, M.~Faizal, A.~F.~Ali and F.~Hammad,
  ``Branes in Gravity's Rainbow,''
  Eur.\ Phys.\ J.\ C {\bf 76}, 264 (2016)
%  doi:10.1140/epjc/s10052-016-4124-7
  [arXiv:1602.04926 [hep-th]].
  %%CITATION = doi:10.1140/epjc/s10052-016-4124-7;%%
  %1 citations counted in INSPIRE as of 18 May 2016

%67
%\cite{Tao:2016baz}
\bibitem{Tao:2016baz}
  J.~Tao, P.~Wang and H.~Yang,
  ``Free-fall frame black hole in gravity¡¯s rainbow,''
  Phys.\ Rev.\ D {\bf 94}, 064068 (2016)
  %doi:10.1103/PhysRevD.94.064068
  [arXiv:1602.08686 [gr-qc]].
  %%CITATION = doi:10.1103/PhysRevD.94.064068;%%
  %2 citations counted in INSPIRE as of 06 Sep 2017

%68
%\cite{Yadav:2016nfh}
\bibitem{Yadav:2016nfh}
  G.~Yadav, B.~Komal and B.~R.~Majhi,
  ``Rainbow Rindler metric and Unruh effect,''
  arXiv:1605.01499 [gr-qc].
  %%CITATION = ARXIV:1605.01499;%%
  %4 citations counted in INSPIRE as of 07 Sep 2017

%69
%\cite{Gangopadhyay:2016rpl}
\bibitem{Gangopadhyay:2016rpl}
  S.~Gangopadhyay and A.~Dutta,
  ``Constraints on rainbow gravity functions from black hole thermodynamics,''
  EPL {\bf 115}, 50005 (2016)
  %doi:10.1209/0295-5075/115/50005
  [arXiv:1606.08295 [gr-qc]].
  %%CITATION = doi:10.1209/0295-5075/115/50005;%%
  %6 citations counted in INSPIRE as of 06 Sep 2017


%70
%\cite{Hendi:2016njy}
\bibitem{Hendi:2016njy}
  S.~H.~Hendi, S.~Panahiyan, B.~Eslam Panah, M.~Faizal and M.~Momennia,
  ``Critical behavior of charged black holes in Gauss-Bonnet gravity¡¯s rainbow,''
  Phys.\ Rev.\ D {\bf 94}, 024028 (2016)
  %doi:10.1103/PhysRevD.94.024028
  [arXiv:1607.06663 [gr-qc]].
  %%CITATION = doi:10.1103/PhysRevD.94.024028;%%
  %27 citations counted in INSPIRE as of 06 Sep 2017

%71
%\cite{Banerjee:2016zcu}
\bibitem{Banerjee:2016zcu}
  R.~Banerjee and R.~Biswas,
  ``Thermodynamics of black holes in rainbow gravity,''
  arXiv:1610.08090 [gr-qc].
  %%CITATION = ARXIV:1610.08090;%%
  %2 citations counted in INSPIRE as of 06 Sep 2017

%72
%\cite{Hendi:2016hbe}
\bibitem{Hendi:2016hbe}
  S.~H.~Hendi, S.~Panahiyan, S.~Upadhyay and B.~Eslam Panah,
  ``Charged BTZ black holes in the context of massive gravity¡¯s rainbow,''
  Phys.\ Rev.\ D {\bf 95}, 084036 (2017)
  %doi:10.1103/PhysRevD.95.084036
  [arXiv:1611.02937 [hep-th]].
  %%CITATION = doi:10.1103/PhysRevD.95.084036;%%
  %15 citations counted in INSPIRE as of 06 Sep 2017

%73
%\cite{Momeni:2017cvl}
\bibitem{Momeni:2017cvl}
  D.~Momeni, S.~Upadhyay, Y.~Myrzakulov and R.~Myrzakulov,
  ``Cosmic string in gravity¡¯s rainbow,''
  Astrophys.\ Space Sci.\  {\bf 362}, 148 (2017)
  %doi:10.1007/s10509-017-3138-4
  [arXiv:1703.00222 [gr-qc]].
  %%CITATION = doi:10.1007/s10509-017-3138-4;%%

%74
%\cite{Alsaleh:2017oae}
\bibitem{Alsaleh:2017oae}
  S.~Alsaleh,
  ``Thermodynamics of rotating Kaluza-Klein black holes in gravity¡¯s rainbow,''
  Eur.\ Phys.\ J.\ Plus {\bf 132}, 181 (2017)
  %doi:10.1140/epjp/i2017-11501-2
  [arXiv:1704.07404 [gr-qc]].
  %%CITATION = doi:10.1140/epjp/i2017-11501-2;%%
  %1 citations counted in INSPIRE as of 06 Sep 2017

%75
\bibitem{Feng:1708}
 Z.-W. Feng and S.-Z. Yang,
  ``Thermodynamics phase transition of a bleck hole in rainbow gravity,''
  [arXiv:1708.06627].

%76
%\cite{Deng:2017umx}
\bibitem{Deng:2017umx}
  X.~M.~Deng and Y.~Xie,
  ``Gravitational time advancement under gravity's rainbow,''
  Phys.\ Lett.\ B {\bf 772}, 152 (2017).
  %doi:10.1016/j.physletb.2017.06.036
  %%CITATION = doi:10.1016/j.physletb.2017.06.036;%%


%%%%%%%%%%%%%%%%%%%%%%%%%%%%%%%%%%%%%%%%%%%%%%%%%%%%%%%%%%%%%5
% in the introduction; the text
%%%%%%%%%%%%%%%%%%%%%%%%%%%%%%%%%%%%%%%%%%%%%%%%%%%%%%%%%%%%%


%77
%\cite{Ali:2014xqa}
\bibitem{Ali:2014xqa}
  A.~F.~Ali,
  ``Black hole remnant from gravity¡¯s rainbow,''
  Phys.\ Rev.\ D {\bf 89}, 104040 (2014)
%  doi:10.1103/PhysRevD.89.104040
  [arXiv:1402.5320 [hep-th]].
  %%CITATION = doi:10.1103/PhysRevD.89.104040;%%
  %45 citations counted in INSPIRE as of 15 Mar 2017

%78
%\cite{Gim:2014ira}
\bibitem{Gim:2014ira}
  Y.~Gim and W.~Kim,
  ``Thermodynamic phase transition in the rainbow Schwarzschild black hole,''
  JCAP {\bf 1410}, 003 (2014)
%  doi:10.1088/1475-7516/2014/10/003
  [arXiv:1406.6475 [gr-qc]].
  %%CITATION = doi:10.1088/1475-7516/2014/10/003;%%
  %19 citations counted in INSPIRE as of 15 Mar 2017

%79
%\cite{Kim:2016qtp}
\bibitem{Kim:2016qtp}
  Y.~W.~Kim, S.~K.~Kim and Y.~J.~Park,
  ``Thermodynamic stability of modified Schwarzschild-AdS black hole in rainbow gravity,''
  Eur.\ Phys.\ J.\ C {\bf 76}, 557 (2016)
  % doi:10.1140/epjc/s10052-016-4393-1
  [arXiv:1607.06185 [gr-qc]].
  %%CITATION = doi:10.1140/epjc/s10052-016-4393-1;%%
  %4 citations counted in INSPIRE as of 24 Aug 2017


%80
%\cite{AmelinoCamelia:2008qg}
\bibitem{AmelinoCamelia:2008qg}
  G.~Amelino-Camelia,
  ``Quantum-Spacetime Phenomenology,''
  Living Rev.\ Rel.\  {\bf 16}, 5 (2013)
  %doi:10.12942/lrr-2013-5
  [arXiv:0806.0339 [gr-qc]].
  %%CITATION = doi:10.12942/lrr-2013-5;%%
  %300 citations counted in INSPIRE as of 12 Sep 2017

%81
%\cite{Lukierski:1993wx}
\bibitem{Lukierski:1993wx}
  J.~Lukierski, H.~Ruegg and W.~J.~Zakrzewski,
  ``Classical quantum mechanics of free kappa relativistic systems,''
  Annals Phys.\  {\bf 243}, 90 (1995).
  doi:10.1006/aphy.1995.1092
  [hep-th/9312153].
  %%CITATION = doi:10.1006/aphy.1995.1092;%%
  %358 citations counted in INSPIRE as of 01 Mar 2016


\end{thebibliography}
\end{document}